\documentclass{ws-ijmpd}
\begin{document}
\markboth{D. V. Ahluwalia-Khalilova}{Minimal spatio-temporal extent of 
events}
%............................ definitions ...............
\def\beq{\begin{eqnarray}}
\def\eeq{\end{eqnarray}}
\def\defn{\buildrel \rm def \over =}
\def\pre{\textit{Preprint~}}

%....................................... Various boldface symbols
\def\s{\mbox{\boldmath$\displaystyle\mathbf{\sigma}$}}
\def\bp{\mbox{\boldmath$\displaystyle\mathbf{\pi}$}}
\def\be{\mbox{\boldmath$\displaystyle\mathbf{\eta}$}}
\def\J{\mbox{\boldmath$\displaystyle\mathbf{J}$}}
\def\K{\mbox{\boldmath$\displaystyle\mathbf{K}$}}
\def\P{\mbox{\boldmath$\displaystyle\mathbf{P}$}}
\def\p{\mbox{\boldmath$\displaystyle\mathbf{p}$}}
\def\hp{\mbox{\boldmath$\displaystyle\mathbf{\widehat{\p}}$}}
\def\x{\mbox{\boldmath$\displaystyle\mathbf{x}$}}
\def\X{\mbox{\boldmath$\displaystyle\mathbf{X}$}}
\def\0{\mbox{\boldmath$\displaystyle\mathbf{0}$}}
\def\bv{\mbox{\boldmath$\displaystyle\mathbf{\varphi}$}}
\def\bx{\mbox{\boldmath$\displaystyle\mathbf{\xi}$}}
\def\bs{\mbox{\boldmath$\displaystyle\mathbf{\sigma}$}}
\def\bc{\mbox{\boldmath$\displaystyle\mathbf{\chi}$}}

\def\hbv{\mbox{\boldmath$\displaystyle\mathbf{\widehat\varphi}$}}
\def\hbxi{\mbox{\boldmath$\displaystyle\mathbf{\widehat\xi}$}}
\def\bn{\mbox{\boldmath$\displaystyle\mathbf{\nabla}$}}
\def\bl{\mbox{\boldmath$\displaystyle\mathbf{\lambda}$}}
\def\br{\mbox{\boldmath$\displaystyle\mathbf{\rho}$}}
\def\1{1}
\def\ar{\stackrel{\hspace{0.04truecm}grav. }{\mbox{$\longrightarrow$}}}
%......................................................................

\title{MINIMAL SPATIO-TEMPORAL EXTENT OF EVENTS, NEUTRINOS,
AND THE COSMOLOGICAL CONSTANT PROBLEM\footnote{This essay
received an ``honorable mention'' in the 2005 Essay
Competition of the Gravity Research Foundation.}}

\author{D. V. AHLUWALIA-KHALILOVA}

\address{Ashram  for the Studies of the Glass Bead Game (ASGBG)\\
Ap. Postal C 600, Zacatecas, ZAC 98060, Mexico\\
and\\
Center for Mathematical, Physical, and Biological Structure 
of the Universe (CIU)\\
Department of Mathematics, University of Zacatecas (UAZ)\\
 Zacatecas, ZAC 98060, Mexico\\
d.v.ahluwalia-khalilova@heritage.reduaz.mx\\}

\maketitle

\begin{abstract} 
Chryssomalakos and Okon, through a
uniqueness analysis, have strengthened the Vilela Mendes suggestion
that the immunity to infinitesimal perturbations in the structure
constants of a physically-relevant Lie algebra should be raised to the
status of a physical principle. Since the Poincar\'e-Heisenberg
algebra does not carry the indicated immunity it is suggested that the
Lie algebra for the interface of the gravitational and quantum realms
(\texttt{IGQR}) is its stabilized form.  It carries three additional
parameters: a length scale pertaining to the Planck/unification scale,
a second length scale associated with cosmos, and a new dimensionless
constant.  Here, I show that the adoption of the stabilized
Poincar\'e-Heisenberg algebra (\texttt{SPHA})  for the \texttt{IGQR} 
has the immediate
implication that `point particle' ceases to be a viable physical
notion. It must be replaced by objects which carry a well-defined,
representation space dependent, minimal spatio-temporal extent.  The
ensuing implications have the potential, without spoiling any of the
successes of the standard model of particle physics, to resolve the
cosmological constant problem while concurrently offering a
first-principle hint as to why there exists a coincidence between
cosmic vacuum energy density and neutrino masses.
The main theses which the essay presents is the following:
an extension of the present-day physics 
to a framework which respects \texttt{SPHA} should be seen as the
most natural and systematic path towards gaining a deeper
understanding of outstanding questions, if not providing answers
to them. 
\end{abstract}


\section{Two problems and a coincidence} 

Poincar\'e and Heisenberg algebras, supplemented by
the principle of local gauge invariance,
play a pivotal and defining role in the formulation and foundations of 
modern physics. If I do not invoke equivalence principle directly
it is because, following Steven Weinberg, I take the view
that Poincar\'e spacetime symmetries, in conjunction with
Heisenberg algebra, not only define 
the notion of point particle but also suggest the equality
of the inertial and gravitational masses. 
\cite{Weinberg:1964ew}\cdash\cite{Ahluwalia-Khalilova:2005jn}

Seen in this light,
irrespective of 
one's preference
for a candidate theory
of quantum gravity, or, the theory of everything,\footnote{See, e.g.,
Refs.~\refcite{'tHooft:2005ym}--\refcite{Adunas:2000zm}
%\cite{'tHooft:2005ym,Rovelli:2003wd,Polchinski:1998Joseph,Rovelli:2004Carlo,
% Amelino-Camelia:2005qa,Corichi:2005fw,Collins:2004bp,Ashtekar:2004vs,Horowitz:2004rn,
% Bojowald:2004zf,Kiefer:2004hv,Sarkar:2002mg,Klauder:2001bp,Carlip:2001wq,Alfaro:1999wd,
% Harbach:2005yu,Grumiller:2004yq,Adunas:2000zm}
for a wide 
spectrum of views and discussions. 
}
a general question that may be asked is 
as to what extent the Poincar\'e and Heisenberg
algebras remain undeformed at the
interface of gravitational and quantum realms (\texttt{IGQR}). 
Further, as a secondary question, as to what
extent the notion of point particle becomes
untenable in any such deformation.

The answer to these questions has a direct impact
on the conceptual and internally consistent formulation
of any quantum theory of gravity.
For instance,
a
consistent Lorentz-covariant quantum theory of strings
requires not only gravity but it carries the advantage
that short distance divergences
of the field theory no longer exist.
While this is far from a trivial justification
in favor of abandoning the notion of point particle,\cite{Polchinski:1994mb}
it would be desirable to provide a deeper first-principle 
reason for doing so. In addition, it may have direct implication on how
theories of extended objects are formulated, or how other 
quantum gravity programs
are implemented. 

The setting thus presented cannot but leave the reader with the 
expectation that perhaps an argument is now to be presented
that Poincar\'e and Heisenberg algebras must suffer a deformation
at \texttt{IGQR}. Indeed that is the case. But, as I hope
to argue below, the deformation
arises not as an ad hoc suggestion but is based on a principle
that requires Lie algebraic stability as a minimal 
requirement for any physically-viable
algebra.

Concurrently with this circumstance there exists
yet another problem which challenges the underlying
Poincar\'e-Heisenberg algebraic structure of the 
standard quantum field theory.
The latter predicts that each quantum field carries
a non-vanishing zero point energy and that
for the standard model fields these quantum fluctuations
do not cancel.  This
result is deeply intertwined with the underlying 
Poincar\'e-Heisenberg algebra.

To make the statement of the problem explicit,  
recall and consider\cite{Weinberg:1996xe}
an effective field theory that takes into account only degrees of
freedom with energies below about $100\;\mbox{GeV}$, with all higher 
energy radiative corrections buried in corrections to various 
parameters in the effective Lagrangian. In  this effective field
theory, the all-pervading cosmic 
vacuum energy density, that serves 
to explain the recently inferred cosmic acceleration, 
\cite{Riess:1998cb,Perlmutter:1998np}
may be symbolically written as
\beq
\rho_{vac} = \frac{1}{2}\sum \hbar\omega = \frac{ c^4 \Lambda}{8 \pi G}
\eeq
where $\Lambda$ is the cosmological constant and the sum  symbolizes,
(a)
contribution of all zero point energies  in the fields of the effective field
theory (with due regard for the sign for fermionic and
bosonic fields), and (b) it is cut off at particle energies equal 
to roughly $100\;\mbox{GeV}$. In units with $\hbar =c=1$, we have
\beq
\frac{1}{2}\sum \hbar\omega\approx \left(100\;\mbox{GeV}\right)^4\,.
\eeq
However, observations do not allow $\rho_{vac}$ to be significantly
greater than the critical density $10^{-48}\;\mbox{GeV}^4$; or,
equivalently $(10^{-3} \;\mbox{eV})^4$. 
This mismatch of fifty-six orders of magnitude (in these units)
is called the \textit{cosmological constant problem} and may be
interpreted as an outstanding failure of the Poincar\'e-Heisenberg
Lie algebraic structure of the standard model of particle 
physics.
It goes without saying that any well-motivated 
deformation of the Poincar\'e-Heisenberg algebra may have
a direct impact on the cosmological constant problem.

This situation is made even more intriguing by the fact
that  there exists a coincidence between the observed 
cosmic vacuum energy density $\rho_{vac}$ and the neutrino masses.\cite{Hung:2003jb,Kaplan:2004dq}

\subsection{Yet, why is the standard model so successful?}

That the discordance manifests itself
so dramatically in an otherwise successful model of particle
physics thus becomes a problem in itself.
Within the context of this essay the answer 
resides in the
fact that the Lie algebraic stability leaves the Lorentz sector
intact, while introducing modifications via length scales which are
either unaccessible at low energies, or carry cosmological
dimensions. 

\section{Introducing the principle of Lie-algebraic stability}

First, in all successful physical theories
Lie algebras have  played a pivotal role in defining the
fundamental notion of \textit{particle} and its evolution.
Secondly, as emphasized by  Chryssomalakos,\cite{Chryssomalakos:2001nd}
Lie algebras naturally divide themselves in two classes.
Those which are \textit{stable} and those which are \textit{unstable}.
Under infinitesimal perturbations  in their structure constants,
the former are isomorphic to all Lie algebras in their vicinity,
while the latter are not. 

Following Vilela Mendes\cite{VilelaMendes:1994zg} 
and Chryssomalakos, \cite{Chryssomalakos:2001nd}
I here suggest that the immunity to  infinitesimal perturbations
in the structure constants 
of a physically-relevant Lie algebra should be raised to the status 
of a principle.\footnote{Furthermore,
this should be considered as a minimal algebraic requirement, and not 
as the most general one.}
 
The reason for this suggestion
lies in the fact that, in retrospect,
the quantum and  relativistic revolutions
can be seen as to have been born from
a unifying Lie-algebraic stability theme.
 That is, quantum and relativistic frameworks
correspond to the Lie-algebraic
stabilization of the algebras underlying the classical mechanics and
Galilean relativity. By promoting the Lie algebraic stability to
a principle, side by side,  say, the principle of
local gauge covariance, one hopes that a bewildering set of 
possibilities that a theorist encounters can be further 
narrowed. I hasten to add that this is not  an abstract idea, but
a paradigm which relies on physical and mathematical robustness
of the underlying algebraic structures. That only such robust frameworks
have a chance of describing physical reality follows if one one wishes
to avoid various fine tuning problems which can ultimately, and often, 
be traced back to stability versus instability of the underlying
Lie algebras. Stability, I then conjecture,
equates to absence of a fine tuning  in a physical theory.

\section{Stabilized  Poincar\'e-Heisenberg algebra} 

In the standard general relativistic and quantum framework,
a freely falling frame at the \texttt{IGQR}  carries 
the Poincar\'e-Heisenberg algebra. Within this framework
while the position and momentum derive their operational
meaning from the fundamental commutator $[x,p_x] = i \hbar$, $\ldots$,
the \textit{vanishing} 
commutators for the operators associated with position allow an 
uncertainty-free ($\Delta x \Delta y =0\;\mbox{etc.}$)
specification of an event. This underlies the operational framework
for the notion of a point particle. 

At the same time, as is apparent
from the work of Wigner, see, e.g.,  Ref.~\refcite{Wigner:1939cj}, 
the Poincar\'e
spacetime symmetries provide kinematical wave equations which describe
the world lines of these point particles. The quantum aspect is
then implemented by 
\begin{enumerate}

\item[\textemdash~]
Using the kinematical wave equation to define a Lagrangian density,

\item[\textemdash~]
Interpreting the functions on which the wave operators act as
field operators, and by imposing Heisenberg's
fundamental commutators/anticommu\-ta\-tors 
for bosonic/fermionic  fields and the Lagrangian-density implied
canonical momenta, and 

\item[\textemdash~] Introducing interactions
by invoking form covariance of the Lagrangian density
under a suitable spacetime-dependent phase transformations 
of the involved fields.
The simplest of these being  a local, i.e. spacetime-dependent,
$U(1)$ transformation
which introduces a massless vector field and results
in quantum electrodynamics.

\end{enumerate}

At the classical level
the effects of a background gravitational field are then incorporated
by demanding form covariance of these wave equations under
general co-ordinate transformations. 
 In making these transformations
the flat spacetime
metric is replaced by appropriate metric compatible with
energy-momentum density associated with the gravitational
background.\footnote{This is where Einstein's equations come
into play.} If one now studies a weak-field and non-relativistic
limit of these wave equations (with a background gravitational
field), and finally invokes Ehrenfest limit, then one verifies that
inertial and gravitational masses indeed cancel out from 
equations of motion. Otherwise, the mass-dependence of test particle
survives in the wave equations and results in either gravitationally-induced
Bohm-Aharonov like effects, or gravitational redshift of flavor oscillation
clocks for neutrinos.\cite{Ahluwalia:1996ev}\cdash\cite{Huang:2005et}

%\cite{Ahluwalia:1996ev,Ahluwalia:1998jx,Konno:1998kq,Wudka:2000rf,Huang:2005et}. 

While these  latter effects may be seen as an implication
of the equivalence principle at \texttt{IGQR} beyond
its original ``$m_i=m_g$'' formulation, the historically-assumed 
mass-independence of the equations of motion no longer survives.
This circumstance, 
for the case of neutron interferometry, was verified
in the classic 1975 experiment of
Colella, Overhauser, and Werner;\cite{Colella:1975dq}
and it continues to inspire similar experiments to
probe \texttt{IGQR}\cite{Chu:1999Steven,Nesvizhevsky:2004qb}.\footnote{It
is noteworthy that in 1997 the pioneering
Werner group 
published a 
discrepancy between the theoretically predicted and experimentally 
measured values of the gravitationally-induced phase
 shift in neutron interferometry  at $1$ part in $10^3$
level.\cite{Littrel:1997kcx}
This discrepancy, to the best of my knowledge, remains 
unexplained.\cite{Adunas:2000zm,Ahluwalia:1999aj}
The fact that no similar signal is seen for the violation
of equivalence principle in atomic interferometry by the Stanford group
of Chu,\cite{Chu:1999Steven} raises the possibility that \textemdash~
despite the unexpectedly large violation of the equivalence principle 
\textemdash~ this may be a quantum gravity effect which manifests itself
only for polarized particles. Such a possibility
naturally occurs in the proposal  of Corichi and Sudarsky.\cite{Corichi:2005fw}  
If the Corichi and Sudarsky's phenomenological 
proposal is indeed at the origin of the
unexpected discrepancy (provided one is
able to reconcile the unexpectedly large 
violation of the equivalence principle), then apart from
offering a possibility for new laboratory experiments
in quantum gravity,
 it may have important
physical  consequences for neutron stars. 
}

While the just summarized observations speak of the great 
strength quantum and relativistic frameworks embody at
\texttt{IGQR}, troubles arise not only at well-known 
attempts to quantize gravity but also in the following two
facts: (a) When \textit{gedanken} experiments
incorporate  gravitational effects into position measurements
the operators associated with the latter cease to 
commute\footnote{This assertion is as valid for 
the measurement of different positional components of
the same event, as for position measurements of two
different events.}\cite{Ahluwalia:1993dd,Doplicher:1994zv}; and, (b)
The Poincar\'e-Heisenberg algebra at \texttt{IGQR} 
induces \textit{irremovable} and \textit{intrinsic}
zero-point energy in freely falling frames. This,\footnote{
Even if one momentarily  does not worry about the
associated problem of cosmological constant. }
implies an intrinsic element of curvature, i.e., gravity, in
freely falling frames.\cite{Ahluwalia-Khalilova:2005jn}
These observations become even more intriguing when, in the
absence of gravity, Sivasubramanian \textit{et al.}\cite{Sivasubramanian:2003nr} 
arrive at non-commutative
geometry for position  measurements of polarized photons.

Within the framework of the paradigm proposed in this
essay, 
the above remarks suggest to question validity of the
Poincar\'e-Heisenberg algebra at \texttt{IGQR}. 
Specifically, the paradigm of Lie algebraic stability suggests
that the problem of constructing
a theory of quantum gravity may lie in the fact that Poincar\'e and Heisenberg
algebras cease to be adequate enough at \texttt{IGQR}. 
Recent physics literature contains numerous efforts to
attend to this suspicion. However, with the exception of the 1994 work
of Vilela Mendes\cite{VilelaMendes:1994zg} (and a few
important works  cited therein) 
essentially all the attempts fail to arise from some deeper
universal principle. The way Vilela Mendes avoids the \textit{ad hoc}
element in his proposal 
is to discover, and point out, that\footnote{Rephrasing the
earlier-noted definition: From a physicists 
point of view a Lie algebra is considered \textit{stable} (or, rigid)
if infinitesimal perturbations in its structure constants results 
in isomorphic algebras. See, e.g., Ref.~ \refcite{Chryssomalakos:2001nd}.}

\begin{enumerate}

\item[\textemdash~] Conceptually, the quantum and relativistic revolutions
of the twentieth century can be viewed as  Lie-algebraic
stabilization of the algebras underlying the classical mechanics and
Galilean relativity. Modulo minor technical remarks, the Poincar\'e
and Heisenberg algebras, \textit{separately}, are endowed with 
Lie algebraic  stability. It was first realized by Faddeev.  
\cite{Faddeev1989}

\item[\textemdash~]
The \textit{combined} Poincar\'e-Heisenberg algebra lacks Lie 
algebraic stability.

\end{enumerate}
Having done that, Vilela Mendes then proceeded to present a stabilized
form of the Poincar\'e-Heisenberg algebra. The \textit{uniqueness} of the 
Vilela Mendes' proposal, with additional elements and  insights, was demonstrated
in the Winter of 2004 by Chryssomalakos and Okon.\cite{Chryssomalakos:2004gk}
This circumstance raised the Lie-algebraic stability from a suggestion
to a new testable principle.

The stabilized Poincar\'e-Heisenberg algebra (\texttt{SPHA}) reads:
\beq
&& \left[J_{\mu\nu},J_{\rho\sigma}\right] = 
i \left(
 \eta_{\nu\rho} J_{\mu\sigma}+\eta_{\mu\sigma} J_{\nu\rho}  
 - \eta_{\mu\rho} J_{\nu\sigma}  
- \eta_{\nu\sigma} J_{\mu\rho} \right)\,,\label{eq:a1}\\
&& \left[J_{\mu\nu}, P_\lambda\right] = i \left(\eta_{\nu\lambda} P_\mu 
- \eta_{\mu\lambda} P_\nu \right)\,,\label{eq:a1b}\\
&& \left[J_{\mu\nu}, X_\lambda\right] = i \left(\eta_{\nu\lambda} X_\mu 
- \eta_{\mu\lambda} X_\nu \right)\,,\\ \nonumber\label{eq:a1c}\\
&& \left[P_\mu,P_\nu\right] = 
 i \left(\frac{\hbar^2}{\ell^2_C}\right) J_{\mu\nu}\,,\\
&& \left[X_\mu,X_\nu\right] =  i {\ell^2_U} J_{\mu\nu}\,, \label{eq:ncst}\\
&& \left[P_\mu,X_\nu\right] = 
i  \hbar \left(\eta_{\mu\nu} \mathcal{F}  +   \beta\, 
J_{\mu\nu}\right)\,,\label{eq:hfc}\\ \nonumber\\
&& \left[P_\mu,\mathcal{F}\right] =
 i \left(\left(\frac{\hbar}{\ell^2_C}\right) X_{\mu} -  \beta P_\mu\right) \,,\\
&& \left[X_\mu,\mathcal{F}\right] = i\left( \beta X_\mu 
- \left(\frac{\ell^2_U}{\hbar}\right) P_{\mu}\right)
\,,\\
&& \left[J_{\mu\nu},\mathcal{F}\right] = 0\,.\label{eq:a2}
\eeq 
Here, $J_{\mu\nu}$ are generators of rotation $\J$ and boosts $\K$ 
($J_{ij}= - J_{ji} = \epsilon_{ijk} J_k$ and $ J_{i0}= - J_{0i} = - K_i$;
Latin indices run over $1,2,3$). $P^\mu$ are generators of spacetime
translations, while $\eta_{\mu\nu} = \mbox{diag}(1,-1,-1,-1)$.
The are several novel features in the stabilized algebra.
First is the existence of two new length scales. One of these  may be 
identified with 
the gravitational unification scale
 $\ell_U \defn \gamma \ell_P$. Here $\ell_P\defn 
\hbar/(m_P\, c) = \sqrt{\hbar G/c^3}$, while $\gamma$ may lie anywhere in the 
range $\10^{-17} \le \gamma \le 1 $, and $m_P$ is the Planck mass.
The other length scale can be taken as  
$\ell_C = \sqrt{  c^4/8 \pi G \rho_{vac}} \defn \sqrt{1/\Lambda}$, 
with $\Lambda$ being the cosmological constant,
and $ \rho_{vac}$ is the vacuum energy density presumably arising
from  the (modified)zero-point energies of various field. In the process,
the underlying algebra unifies the extreme microscopic 
(i.e, Planck/Unification realm) and the extreme 
macroscopic (i.e. cosmological scale). For reasons underling these
specifications, see Refs.~\refcite{Ahluwalia-Khalilova:2005jn}--\refcite{VilelaMendes:1994zg}.\footnote{The analysis presented
in Ref.~\refcite{Sivasubramanian:2003nr}, though carried
out in an entirely different context, when taken to its logical
conclusion suggests that $\ell_U$ may in fact be much larger
than that which appears ``natural.'' That is,  $\gamma$ may be 
significantly less than $10^{-17}$ and it need not be identified
with a unification scale. 
In addition, it may depend on the
particle species which probe the spacetime non-commutativity.
For instance, spacetime as seen by a photon, and as that
probed by a graviton, may
not coincide (except perhaps in some averaged sense).}
Second, there exists a 
new dimensionless constant $\beta\ne 0, \in {R}$.\footnote{Its 
presence has been noted in  Refs.
~\refcite{VilelaMendes:1994zg}--\refcite{Khruschev:2002cq}
with differing emphasis.} Third,
existence of $\mathcal{F}$ which ceases to be central.

For the purposes of this essay, and to hint at the conceptual and predictive
strength of  the principle of  Lie algebraic stability, I now discuss
two issues. The first one concerns the notion of point particle (and
associated questions and observations),
while the second is a preliminary study on the implications for the
cosmological constant problem.

\section{Inevitability of abandoning the notion of point particle,
and related observations}

The fact that the Heisenberg's fundamental
commutator (\ref{eq:hfc}) undergoes  non-trivial modifications
with $\mathcal{F}$ ceasing to be central, and $\beta \ne 0$, has the
following
immediately identifiable 
consequence:
the position-momentum Heisenberg uncertainty 
relations get modified. For example,
\beq
\Delta x \,\Delta p_x  \ge \frac{\hbar}{2}
\left\vert \left\langle 
  \mathcal{F} 
\right\rangle\right\vert\,,\label{eq:xpx}
\eeq
 while $\Delta x\, \Delta p_y $ no longer vanishes, but instead is
given by
\beq
\Delta x\, \Delta p_y  \ge \frac{\beta\hbar}{2}
\left\vert \left\langle 
  J_z 
\right\rangle\right\vert\,.\label{eq:zimpokA}
\eeq
That is, $\Delta x \,\Delta p_x$ is sensitive
to $\mathcal{F}$; while sensitivity to $\beta$ is
carried in $\Delta x\, \Delta p_y$.
Furthermore, in the usual notation, one has the following 
representative expression for the product of uncertainties
in position measurements:
\beq
\Delta x\, \Delta y  \ge \frac{\ell_U^2}{2}
\left\vert \left\langle 
  J_z 
\right\rangle\right\vert\,,\label{eq:xy}
\eeq
with
\beq
\Delta p_x\, \Delta p_y  \ge \frac{\hbar^2}{2\ell_C^2}
\left\vert \left\langle 
  J_z 
\right\rangle\right\vert\,.\label{eq:zimpokB}
\eeq
complementing equation (\ref{eq:xy}) for momentum measurements. 
The expectation value, denoted by $\langle\ldots\rangle$
in the above expressions, is with respect states that arise
in a (yet to be fully formulated) quantum field theory based on
Lie algebra for \texttt{IGQR}, i.e., the \texttt{SPHA}.

Above modified uncertainty relations are to be further supplemented by
relations of the form
\beq
\Delta x \,\Delta t \ge
\frac{\ell_U^2}{2 c}
\left\vert \left\langle 
  K_x 
\right\rangle\right\vert\,.
\eeq
For $e^\pm$ it takes the form\footnote{In equations (\ref{eq:xta})  
below the factor of $i$ is left to remind the reader that
for $(1/2,0)$ Weyl spinor the boost generator is $ - i\s/2$,
while for a  $(0,1/2)$ Weyl spinor the boost generator is $ + i\s/2$.}
\beq
\Delta x \,\Delta t \vert_{e^\pm} \ge
\frac{\ell_U^2}{4 c}
\left\vert \left\langle 
 \left( \begin{array}{cc}
- i \sigma_x & 0 \\
0 & i \sigma_x
  \end{array}\right)
\right\rangle\right\vert\,.\label{eq:xta}
\eeq
For the standard-model $\nu_e$ and ${\overline{\nu}_e}$, the counterparts are
\beq
\Delta x \,\Delta t \vert_{\nu_e} \ge
\frac{\ell_U^2}{4 c}
\left\vert \left\langle 
 \left( \begin{array}{cc}
- i \sigma_x & 0 \\
0 & 0
  \end{array}\right)
\right\rangle\right\vert,\;
 \Delta x \,\Delta t \vert_{\overline{\nu}_e} \ge
\frac{\ell_U^2}{4 c}
\left\vert \left\langle 
 \left( \begin{array}{cc}
0 & 0 \\
0 &  i \sigma_x
  \end{array}\right)
\right\rangle\right\vert\,.\label{eq:xtbb}
\eeq

For a massive vector particle $B^\mu$, using the results given in
Ref.~ \refcite{Ahluwalia:2000pj} I obtain\footnote{The needed from of
$K_x$ is obtained from ``$K_x$'' given in Eq. (4) of the indicated
reference, and then evaluating $S$``$K_x$''$S^{-1}$ (where $S$ is
given by Eq. (18) of the said reference).}
\beq
\Delta x \,\Delta t \vert_{B^\mu} \ge
\frac{\ell_U^2}{2c}
\left\vert \left\langle 
 \left( \begin{array}{cc}
- i \sigma_x & 0 \\
0 & 0
  \end{array}\right)
\right\rangle\right\vert\,.\label{eq:xtc}
\eeq

Reader's attention is drawn to different numerical factors in
the right hand sides of Eqs. (\ref{eq:xta})-(\ref{eq:xtc}), and
that states that appear in $\langle\ldots\rangle$ correspond 
to the indicated particles. The species dependence of these
relations is reminiscent of the results found
in Ref.~\refcite{Sivasubramanian:2003nr} (where it is apparent,
though not explicitly stated, that deciphered 
granularity of the spacetime is probe dependent).

For comparison, Eqs. (\ref{eq:zimpokA})-(\ref{eq:zimpokB})
for $e^\pm$  take the form
\beq
&& \Delta x\, \Delta p_y\vert_{e^\pm}  \ge \frac{\beta\hbar}{4}
\left\vert \left\langle 
  \left(
       \begin{array}{cc}
        \sigma_z & 0 \\
        0 & \sigma_z
	\end{array}
  \right)
\right\rangle\right\vert\,
\\
&& \Delta x\, \Delta y\vert_{e^\pm}   \ge \frac{\ell_U^2}{4}
\left\vert \left\langle 
    \left(
       \begin{array}{cc}
        \sigma_z & 0 \\
        0 & \sigma_z
	\end{array}
  \right)
\right\rangle\right\vert\\
&& \Delta p_x\, \Delta p_y\vert_{e^\pm}   \ge \frac{\hbar^2}{4\ell_C^2}
\left\vert \left\langle 
    \left(
       \begin{array}{cc}
        \sigma_z & 0 \\
        0 & \sigma_z
	\end{array}
  \right) 
\right\rangle\right\vert\,.
\eeq
For $\nu_e$ and $\overline{\nu}_e$ of the standard model, 
 Eqs. (\ref{eq:zimpokA})-(\ref{eq:zimpokB}) have the following
explicit form
\beq
&& \hspace{-26pt}\Delta x\, \Delta p_y\vert_{\nu_e}  \ge \frac{\beta\hbar}{4}
\left\vert \left\langle 
  \left(
       \begin{array}{cc}
        \sigma_z & 0 \\
        0 & 0
	\end{array}
  \right)
\right\rangle\right\vert,\;
 \Delta x\, \Delta p_y\vert_{\overline{\nu}_e}  \ge \frac{\beta\hbar}{4}
\left\vert \left\langle 
  \left(
       \begin{array}{cc}
        0 & 0 \\
        0 & \sigma_z
	\end{array}
  \right)
\right\rangle\right\vert\,
\\
&& \hspace{-26pt}\Delta x\, \Delta y\vert_{\nu_e}   \ge \frac{\ell_U^2}{4}
\left\vert \left\langle 
    \left(
       \begin{array}{cc}
        \sigma_z & 0 \\
        0 & 0
	\end{array}
  \right)
\right\rangle\right\vert,\;
\Delta x\, \Delta y\vert_{\overline{\nu}_e}   \ge \frac{\ell_U^2}{4}
\left\vert \left\langle 
    \left(
       \begin{array}{cc}
      0 & 0 \\
        0 &   \sigma_z
	\end{array}
  \right)
\right\rangle\right\vert\,
\\
&& \hspace{-26pt}\Delta p_x\, \Delta p_y\vert_{\nu_e}   
\ge \frac{\hbar^2}{4\ell_C^2}
\left\vert \left\langle 
    \left(
       \begin{array}{cc}
          \sigma_z & 0 \\
        0 &   0
	\end{array}
  \right) 
\right\rangle\right\vert,\;
\Delta p_x\, \Delta p_y\vert_{\overline{\nu}_e}   \ge \frac{\hbar^2}{4\ell_C^2}
\left\vert \left\langle 
    \left(
       \begin{array}{cc}
          0& 0 \\
        0 &   \sigma_z
	\end{array}
  \right)
\right\rangle\right\vert  \,.\label{eq:zimpokD}
\eeq
Whereas for massive vector particles, the counterpart of these
is obtained from 
Eqs. (\ref{eq:zimpokA})-(\ref{eq:zimpokB}) by the replacement
\beq
J_z\vert_{B^\mu} \rightarrow
\left(
\begin{array}{cccc}
0 & 0 & 0 & 0 \\
0 & 0 & -i & 0 \\
0 & i & 0 & 0\\
0 & 0 & 0 & 0 
\end{array}
\right)\,.
\eeq

The spatio-temporal extent, besides representation space, depends
on $\ell^2_U$; the uncertainty products such as $\Delta p_x\, \Delta p_y$
depend on $\hbar^2 \ell_C^{-2}$. As a reminder,
$\Delta x \,\Delta p_x$ is sensitive
to $\mathcal{F}$; while sensitivity to $\beta$ is
carried in $\Delta x\, \Delta p_y$.

Referring to Eqs. (\ref{eq:a1}) and (\ref{eq:a1b}),  note is to be taken 
that since the Lorentz sector remains
intact the $\J^2$ and $J_z$ still commute (while $J_z$ does not
commute with $J_x$ and $J_y$). This allows to choose states with well-defined
$\J^2$ and $J_z$. If we tentatively identify $\J^2$ with the standard model
fermions and bosons, 
then its eigenvalues, with exception of Higgs, are non-zero.
That is, all matter and gauge field (with exception of Higgs) cannot
be identified as point particles. Their position measurements carry
a fundamental and irreducible uncertainty. If in Eq.  (\ref{eq:xy}),
$\Delta y$ is taken as zero these particles acquire the
interpretation of string-like objects. 
Or, if $\Delta y \approx \ell_U$, then
one obtains the interpretation of a membrane-like entity. Yet, 
for physical states for which $\left\langle 
  J_z 
\right\rangle$ vanishes, the point-like interpretation holds.
The fundamental spatial extension is bounded from below by
$(\ell_U^2/{2})
\left\vert \left\langle 
  J_z 
\right\rangle\right\vert$; and it vanishes for a small
subset of states for which 
$ \left\langle 
  J_z\right\rangle$  is zero.

As such point particle ceases to be a viable notion in \texttt{IGQR}.
Furthermore, a concrete modification is suggested for the
algebra underlying freely falling frames. It consist of replacing
the Poincar\'e and Heisenberg algebras by the Lie stabilized
Poincar\'e-Heisenberg algebra, \texttt{SPHA}. The latter governs and defines
the evolution of the  emergent extended objects. 

The following
questions and observations immediately arise and 
may be worthy of systematic exploration: 

\begin{enumerate}

\item  {\it Modification to wave-particle duality \textemdash~}
Since now $p_x \ne \frac{\hbar}{i} \frac{\partial}{\partial x}$, the  
considerations found in Refs.~\refcite{Kempf:1994su} and~\refcite{Ahluwalia:2000iw} suggest
a fundamental modification to the wave-particle duality. That is, de Broglie
relation $\lambda_{dB} = h/p$ is no longer viable and must suffer 
a well-defined
modification. In particular, I expect it to have same qualitative 
behavior as found in Ref.~\refcite{Ahluwalia:2000iw}. That is, the modified
$\lambda_{dB}$ saturates to $\ell_U$ as $p\to\infty$. In fact,
given that spatial co-ordinates of an event no longer commute,
and that point particle is no longer a viable notion, suggests
that an event carrying momentum $\p$ is characterized  by a 
set of wavelengths. 

\item {\it Lack of primitiveness of $X_\mu$  \textemdash~}
What physical interpretation is to be associated with the
non-commutative $X_\mu$. It's interpretation as space-time
coordinates lacks the required primitive nature  for
Lie algebra generators, as noted by
Chryssomalakos and Okon.\cite{Chryssomalakos:2004gk}

\item  {\it Configuration-space wave equations \textemdash~}
If one assumes that the boost parameter, in the notation of
Ref.~\refcite{Ahluwalia-Khalilova:2005jn}, remains 
unchanged\footnote{It is not obvious that such an assumption is valid.
For one things, the notion of inertial frames now requires a careful
examination and spacetime now carries non-commutative elements.
Yet, such an assumption is consistent with dispersion relation
$E^2=p^2+m^2$. But, I see no reason that the dispersion relation itself
should not suffer a modification. 
}
\[
\cosh\varphi = \frac{E}{m},\quad\sinh \varphi = \frac{p}{m},\quad
\hbv=\frac{\p}{p}
\]
then, the 
momentum-space wave equations for the \texttt{SPHA}
remain intact, but their configuration-space form depend on (a) resolution
of the question just enumerated, and (b) the precise conceptual
understanding and form of $P_\mu$ as a differential operator which solves the
\texttt{SPHA}.

\item  {\it   Lagrangian density, and quantization rules \textemdash~}
The answer to the above question guides to write down the quantum field
operator, to obtain the Lagrangian density (note it immediately
follows once ``configuration space'' wave equation is known), 
and to define $J_{\mu\nu}$ and
$P_\mu$, and perhaps $X_\mu$ and $\mathcal{F}$, in terms of the field
operator. The quantization 
rules may then be obtained by demanding that the resulting objects
satisfy the \texttt{SPHA}.

\item  {\it   Discrete symmetries \textemdash~}
How are the notions of charge conjugation, parity, and time reversal
defined. Is the theory symmetric under modified form of these
symmetries? Answer to this question, e.g.,  carries relevance to
the observed matter-antimatter asymmetry in the universe. 

\item
{\it S-matrix \textemdash~}
What is the associated S-matrix structure?

\item {\it Equivalence principle \textemdash~}
Do the notions of inertial and gravitational masses undergo any
change? This question can be 
examined by concurrently studying the non-relativistic and
Ehrenfest limit of the ``configuration space'' wave equations and 
by  repeating the 1964 analysis of Ref.~\refcite{Weinberg:1964ew}
in the new context. 

\end{enumerate}

\section{Impact of \texttt{SPHA} on the
cosmological constant problem}

The cosmological constant problem and the zero point energy for
bosonic and fermionic field are directly related, and rest on
Poincar\'e and Heisenberg algebras. In order to define
the impact of the Lie algebraic stabilization on the cosmological
constant problem it is first helpful to add a few brief comments. These
complement the discussion given in the opening section. The remarks
are then followed by the subject matter of the impact of Lie 
algebraic stabilization of the 
Poincar\'e-Heisenberg algebra on the cosmological constant problem.

\textit{Few brief remarks on the standard zero point energy  \textemdash~}
Within the context of 
Heisenberg algebra, a 
heuristic understanding of the zero point energy is gained by
considering a one-dimensional non-relativistic 
harmonic oscillator. In the standard notation, it is characterized
by the Hamiltonian: $
H= \frac{p_x^2}{2 m} + \frac{1}{2} m\, \omega^2\, x^2$.
The zero point energy of $\frac{1}{2}\hbar\omega$ arises
directly when one determines the eigenspectrum of $H$ 
with $x$ and $p_x$ satisfying the fundamental Heisenberg
commutator $[x,p_x]=i\hbar.$ It corresponds to the energy of the ground
state. In obtaining this result, the 
spacetime is assumed to be commutative. 
In transition from quantum mechanics of point particles, to a relativistic
quantum
field describing point particles, instead of requiring
 $[x,p_x]=i\hbar$  (with  commutative spacetime), 
one now imposes the same relations, with right hand side 
now being a Dirac delta function, $i \hbar \delta(\x-\x^\prime)$, or zero
(for field-field, and momentum-momentum, commutators),
 and $x \rightarrow \psi(x)$, the field,
while $p_x \rightarrow \pi(x)$, with the latter representing
the canonically conjugate momentum associated with $\psi(x)$;
and further replacing the commutator by  anticommuator 
if the field is fermionic. For the standard model fields, 
each of the the fermionic fields, as is well known, is found to carry
a zero point energy of $- \frac{1}{2}
\hbar \omega$, while each of the bosonic fields carries 
 $+ \frac{1}{2}
\hbar \omega$, for each mode of angular frequency $\omega$.
The cosmological constant problem arises because for the 
standard model fields the bosonic and fermionic 
contributions do not cancel, and
because these contributions when summed over all accessible energies,
up to a cut off, give a result which violently disagrees with observational
data.\footnote{The numerical aspect of the disagreement 
was made specific in the opening section of  this essay.}

\subsection{Zero point energy with Lie-algebraically \texttt{SPHA}: 
Naive arguments}

Naively, one may begin with $H$, considered above, 
for the simple harmonic oscillator. 
Such an exercise, with $\beta=0$, has been
undertaken by Vilela Mendes.\cite{VilelaMendes:1999xv}
The result of Vilela Mendes can be put in a closed form
for the ground state if one sets  $n=0$ in Eq.
43 of Ref.~\refcite{VilelaMendes:1999xv}, identify a dimensionless
parameter $\zeta$, notice a pattern in the leading order
terms, and then sum the indicated series. This set of steps
results in a closed form expression for the modified zero
point energy, and reads\footnote{The indicated domain of validity in Eq. (\ref{eq:HPsho}),
corresponds to Vilela Mendes assumption
\[
\frac{\hbar^2}{4 \ell_U^4 m^2 \omega^2} \gg 1.
\]
}
\beq
E_0 = \frac{1}{2}\left(1 - \frac{\zeta^2}{1-\zeta^2}\right) \hbar \omega,\quad
\mbox{for}\;\zeta \ll \frac{1}{2\sqrt{2}}\left(
\frac{\hbar/m c}{\ell_U}\right) \label{eq:HPsho}
\eeq   
where the dimensionless parameter $\zeta$ is defined as
\beq
\zeta \defn \frac{1}{2}\left(\frac{\ell_U}{\hbar/m c}\right)
\sqrt\frac{\hbar \omega}{m c^2}
 = 
 \frac{1}{2}\left(\frac{m}{m_U}\right)
\sqrt\frac{\hbar \omega}{m c^2} \,.
\eeq
Here, $m$ represents the mass of the bosonic oscillator, $m_U$ 
corresponds
to cut off mass scale for the effective theory,
and $\omega$ is the angular frequency of oscillation.

For $\zeta \ll 1$, i.e. at `low' angular frequencies, 
the zero point energy $E_0$ remains close to $\frac{1}{2} \hbar \omega$.
Whereas as $\zeta$ approaches the unification scale, the $E_0$  
\textit{vanishes}
at $\zeta = 1/\sqrt{2} \defn \zeta_c$, while concurrently 
one enters the parameter
space where the validity criterion in Eq. (\ref{eq:HPsho}) is crossed.
In terms of the $\zeta_c$, the domain of validity for 
Eq. (\ref{eq:HPsho}) translates to
\beq
\left( \frac{m}{m_U}\right)^2 
\sqrt{\frac{\hbar \omega}{m c^2}} \ll\zeta_c
\eeq
For $m \ll m_U$, the domain of validity extends to 
$\hbar\omega\sim m c^2$. For $m\sim m_U$, the domain of validity is
severely restricted to $\hbar\omega\ll m c^2$.

The result (\ref{eq:HPsho}) is in  sharp contrast
to the \textit{unstable} form of Poincar\'e-Heisenberg where high angular
frequencies result in increasingly higher contributions to
the zero point energy. 

It is quite clear that the preliminary 
considerations presented here point towards dramatic softening, if not 
the complete resolution, of the cosmological constant problem.
A stronger claim cannot be made because
such an analysis is too naive. For one thing, there is no
reason to believe that the form of $H$ 
remains valid when the underlying spacetime is no longer commutative.
More importantly, one has no unique guiding principle to define
as to what one means by a simple harmonic oscillator for the
latter circumstance.
To bypass this problem, it is advisable that heuristic argument given
here serve only as a motivation to look at the spectrum of free
bosonic and fermionic fields as they exist for  (yet to be developed)quantum
fields based on \texttt{SPHA}.
 
If the naive result  contained in Eq. (\ref{eq:HPsho}) 

\begin{enumerate}

\item[\textemdash~] is essentially confirmed by a rigorous
analysis of the quantum fields based on 
\texttt{SPHA}, and if it
\item[\textemdash~] captures the essence of the exact result

\end{enumerate}
then for heavy particles,  $\zeta$ approaches $\zeta_c$ 
faster than  that compared with light particles.
Therefore, the dominant contribution comes to the cosmological
constant from the lightest particles in the standard model, i.e.,
the neutrinos of the fermionic sector and the photons of the bosonic sector. 
Furthermore, for massive particles 
this contribution comes not from the angular frequencies
$\omega \sim m c^2/\hbar$, but from the lower spectrum of
angular frequencies.
It
may underlie the observation that the vacuum energy density associated with
$\Lambda$ is of the same order as that of neutrinos.
Specifically, the coincidence between $\rho_{vac}\approx \left(
10^{-3}\;\mbox{eV}\right)^4$ and neutrino masses as suggested 
by the atmospheric and solar neutrino data, see e.g. Refs.~\refcite{Hung:2003jb} and 
\refcite{Kaplan:2004dq}, 
acquires a plausible first-principle 
explanation in the just outlined scenario.

\section{To sum up}
Within the framework of the standard model of particle physics,
the cosmological constant problem poses
a dramatic discordance between reality 
and prediction. The \texttt{SPHA}, summarized in Eqs.
(\ref{eq:a1})-(\ref{eq:a2}),
 offers 
the next logical step towards extension of the standard model in
\texttt{IGQR}
without spoiling any of its grand successes. In the process it
offers a well-defined departure from the notion of point particle
where an event carries a minimal spatio-temporal extent. The latter
depends on the representation space to which the event belongs.

At this early stage it is difficult to assert with any confidence
if \texttt{SPHA} is indeed the next approximation to the physically-realized 
 algebraic structure
over which to extend the standard model of particle physics (and which
incorporates gravity in its quantum nature). 
However, from a theoretical point of view, 
an extension of the present-day physics 
to a framework which respects \texttt{SPHA} should be seen as 
the most natural and systematic path towards gaining a deeper
understanding of outstanding questions, if not providing answers
to them.

\section*{Acknowledgments} 

It is my pleasure to thank  
Chryssomalis Chryssomalakos,
Daniel Grumiller, and John Swain for their
comments and questions on earlier versions of this essay, and
Daniel Sudarsky for a discussion related to Ref.~\refcite{Corichi:2005fw}.

%\bibliographystyle{JHEP} % this is JHEP style
%\bibliography{DVAK} % this is the bibtex file

\providecommand{\href}[2]{#2}\begingroup\raggedright

\end{document}